\documentstyle[referee,psfig,epsfig]{l-aa}
\begin{document}
\thesaurus{ 06() }
\title{VLBI observations of 3C\,273 at 22\,GHz and 43\,GHz. \\
I: Search for short time-scale structural variation}  

\author{F. Mantovani\inst{1}, W. Junor\inst{2}, C. Valerio\inst{1}, 
I. M$^c$Hardy\inst{3}}
\institute{Istituto di Radioastronomia del CNR, Bologna, Italy
\and Institute for Astrophysics, University of New Mexico, Albuquerque, NM, USA
\and Department of Physics, University of Southampton, UK}
\offprints{fmantovani@ira.bo.cnr.it}
\date{}
\maketitle
\markboth{Mantovani et al.}{Short time-scale structural variation in 3C\,273}
\begin{abstract}
The results of VLBI observations of the quasar 3C\,273,
obtained during a multi-frequency campaign in late 1992 in the radio,
millimeter and X-ray bands are presented and discussed.  The VLBI observations 
were made at 22 GHz with a Global Array and at 43 GHz with the Very Long 
Baseline Array. Hybrid maps and modelfits were made in order to look
for any short time scale structural variations of the inner part of the 
radio jet. In 42 days 3C\,273 was observed 5 times at roughly 10 day intervals. 
The jet structure did not show dramatic changes in that period.
However, we were able to follow the changes in the positions of the components
along the first 2 mas of the jet with respect to the core.  The resolution 
achieved by these observations allows the decomposition of the apparent 
velocity vector of each component into two orthogonal directions. 
We find evidence of reverse motion for a component along the jet.
The jet starts wiggling well inside the first parsec from the core. 
The spatial period of the quasi-sinusoidal oscillating jet path about the 
jet major axis was derived for distances between 2 and 8 mas from the core.   

\end{abstract}
\section{Introduction}
The quasar 3C\,273 (z=0.158; Schmidt 1963) is one of the most studied
superluminal radio sources (e.g. Zensus et al.\ 1990). Due to its brightness
and proximity (1 milliarcsecond corresponds to  1.85 parsec;
$H_{0}$=100\,km\,s$^{-1}\,$Mpc$^{-1}$, $q_{0}$=0.5),
it has been the target of Very Long Baseline Interferometry (VLBI) 
observations over a wide range of radio frequencies.

The radio structure of 3C\,273
shows a well defined core-jet morphology from the mas scale up to the 
arcsecond scale. The jet extends out to $\sim$20
arcseconds with the ridge line of emission showing a clear `wiggle'
(Davis et al.\ 1985). VLBI imaging at frequencies from 5 to
100\,GHz has shown that the ridge line of the jet is curved on a scale
from 0.05 to 25 mas, oscillating around the main orientation of the jet 
(B\aa\aa th et al.\ 1991; Krichbaum et al.\ 1990; Zensus et al.\ 1990).
Zensus et al.\ (1990), examining the positions of the components at 
various epochs, pointed out that major bends 
apparently occur at about 0.2, 3--4, 10--11 and 15--17 mas from the core.
They also suggest that discrepancy in the position angles of some of the
components observed at different frequencies may represent a spectral index 
gradient across the jet which implies a frequency dependence of the jet ridge
line in the evolution of parsec-scale jet. Krichbaum et al.\ (1990) 
investigated the jet structure and showed that there are clear indications
that the jet bends along its entire length down to $\sim$1 mas.  (There are
too few observations interior to this).  They also pointed out that the 
ridge line in the jet has a quasi-sinusoidal shape. 

We present here VLBI observations of 3C\,273 at 22\,GHz and 43\,GHz.
They were performed as part of a multi-frequency campaign
carried out from December 12, 1992 to January 24, 1993. Results from the 
X-ray observations can be found in Leach et al.\ (1995) and for the mm-band
observations in M$^c$Hardy et al.\ (1994). 
Hybrid maps at the two frequencies for four out of five epochs of
observations will be shown. Model-fitting of the final data set obtained
as a result of the imaging/self-calibration process 
had performed also. An analysis of the source structure in the first
8 mas of the jet and measurements for short time-scale structural variations
in 3C\,273 are presented and the results discussed.

\section{Observations and data reduction}

The VLBI observations used an array consisting of the 
VLBA \footnote {Very Long Baseline Array of the National Radio Astronomy 
Observatory, USA} and Effelsberg \footnote {The 100-m telescope of the 
Max Planck Institut f\"ur Radioastronomie, Germany},
Medicina and Noto \footnote {The 32-m twin telescopes of the Istituto di 
Radioastronomia, Italy} at 22\,GHz, while the stand-alone VLBA was used 
at 43\,GHz.   Five sets of observations were made from December 1992 to 
January 1993 at roughly 10 day intervals.  Data were recorded on Mark\,IIIA 
compatible terminals, in Mode E (14 MHz bandwidth).  Typically, each scan had 
lasted 13 minutes. The data were correlated at the Array 
Operations Centre in Socorro, New Mexico. The correlator output was calibrated
 in amplitude and phase using ${\cal AIPS}$ 
\footnote {${\cal AIPS}$ is the NRAO's {\it Astronomical Image Processing 
System}} 
and imaged using DIFMAP \footnote {DIFMAP is part of the 
{\it Caltech VLBI software Package}} (Shepherd et al.\ 1995). 

The amplitude calibration of VLBI observations at 22\,GHz and 43\,GHz, 
is mainly limited by the lack of point source calibrators. System Temperature 
measurements were made at the end of each scan and the gain curve of each 
telescope was used to correct the variation of effective antenna gain 
with antenna elevation for all telescopes.  The calibration accuracy differs 
between observing sessions, owing to different weather conditions.  
(Millimetric observations are sensitive to the water vapor content of the 
atmosphere.)

Images were made for each 'calibrator' source
scheduled for a few scans during the observing program, namely 0804$+$499
and 1611$+$343 at 22\,GHz, OJ\,287 and BL\,Lac at 43\,GHz. 
From an inspection of the visibility plots, it appears that these sources 
were barely resolved with the array and are therefore reasonable calibrators.
The poor {\it uv}-plane coverage ($\sim$3700 and $\sim$5000 
visibilities on average at 22\,GHz and43\,GHz respectively) and the rms noise
($\sim$ 14\,mJy) did not allow these to be mapped properly. Much of
the flux from their jet structure (see Kellermann et al.\ 1998) is 
clearly lost.  We were able to image the central brightest component 
only and most of the emission from the more extended structure is missed.

The self-calibration procedure, which uses closure amplitudes to 
determine telescope amplitude corrections, gave calibration factors that are 
within 10\% of unity.
Total power measurements, contemporary with the VLBI observations, were not 
available for the calibrators.  These sources are known to be highly 
variable with time at high frequencies.  Light curves at
22\,GHz for 1611$+$343 and OJ\,287 can be found in Tornikoski et al.\ (1994).
The ratio between our extrapolated `zero baseline' flux density and the 
total power flux density is $\sim$0.75 for both 1611$+$343 and OJ\,287. 
Tornikoski et al.\ (1994) also report monitoring observations for BL~Lac at 
37\,GHz. Here, the ratio between the two flux densities is $\approx$0.5.  
However, BL~Lac has a more extended structure than 1611$+$343 and OJ\,287
and we were unable to image this successfully. 
 
The total correlated flux density for 3C\,273 at both frequencies is about
one half of the total power measurements with single dish. However, 
correlated flux and total power flux  follow a similar
trend (see, for example von Montigny et al.\ 1997). More recent 22\,GHz VLBA
observations by Lepp\"anen et al.\ (1995) do show a similar ratio. 
In conclusion we estimate that the systematic amplitude calibration errors 
for any of the data sets are $\leq 20\%$.

During session four (14 January 1993), the array was so heavily affected by 
adverse weather conditions that only crude images could be made. 
Data from that session will not be used in the following discussions.
Table 1 summarizes the observational data and imaging. Column 1: project name;
column 2: date of the observation; column 3: observing frequency; column 4: 
stations; 
column 5: beam major axis; column 6: beam minor axis; column 7: major axis
Position Angle; column 8: dynamic range, i.e. the ratio between the peak
flux density and the rms noise in the image, measured far from the source
of emission.
 
\begin{table}[h]
\centerline{\bf Table\,1 - VLBI observations of 3C\,273 }
\vspace{0.5cm}
\hspace{0.5cm} 
\begin{tabular}{llllccrcr}
\hline
Project & Date & Freq. & Stations$^a$ &     & beam &   & rms noise & dynamic \\
      & &  GHz  &              & mas & mas  & deg & mJy/beam     & range \\ 
\hline
bj5a& 12Dec1992& 22.2  & EB,MC,VLBA-HN-SC-FD-KP-LA-NL-PT-BR-OV &1.3&0.2&-8&14&386  \\
    &     & 43    & VLBA-FD-KP-LA-NL-PT-BR-OV             &0.8&0.5& 7&12&393  \\
bj5b& 21Dec1992& 22.2  & EB,MC,VLBA-HN-SC-FD-KP-LA-NL-BR-OV    &1.0&0.2&-8& 5&1266 \\
    &    & 43    & VLBA-HN-FD-KP-LA-NL-PT-BR-OV          &0.7&0.4&-9& 5&1080 \\
bj5c& 04Jan1993& 22.2  & EB,MC,VLBA-HN-SC-FD-KP-LA-NL-BR-OV    &1.0&0.2&-6&10&427 \\
    &    & 43    & VLBA-HN-FD-KP-LA-NL-PT-BR-OV          &0.7&0.4&-5& 5&451  \\
bj5e& 23Jan1993& 22.2  & MC,NT,VLBA-HN-SC-FD-KP-LA-NL-BR       &1.3&0.5& 0& 4&376 \\
    & 24Jan1993& 43    & VLBA-HN-FD-KP-LA-NL-PT-BR             &0.6&0.4&44& 5&566  \\
\hline
\end{tabular}
\vspace{0.5cm}

$^a$: The station labels listed are those for which visibilities were 
      available at the end of the analysis of the raw correlated data  
      for each session, and are as follows:
      EB = Effelsberg (Germany) 100-m, MC = Medicina (Italy) 32-m,
      NT = Noto (Italy) 32-m, VLBA: HN = Hancock, SC = St. Croix,
      FD = Fort Davis, KP = Kitt Peak, LA = Los Alamos, NL = North Liberty,
      PT = Pie Town, BR = Brewster, OV = Owens Valley, (USA) 25-m. Noto did
      not take part in the first two sessions.
\end{table}
\section{The mas structure of 3C\,273}
The images obtained for each epoch at 22\,GHz and 43\,GHz are shown in 
Figs.\,1--8.
The quality of the images is variable  since the data quality from each station,
the calibration accuracy, the ({\it u,v}) coverage change between sessions.
\begin{figure}
\psfig{figure=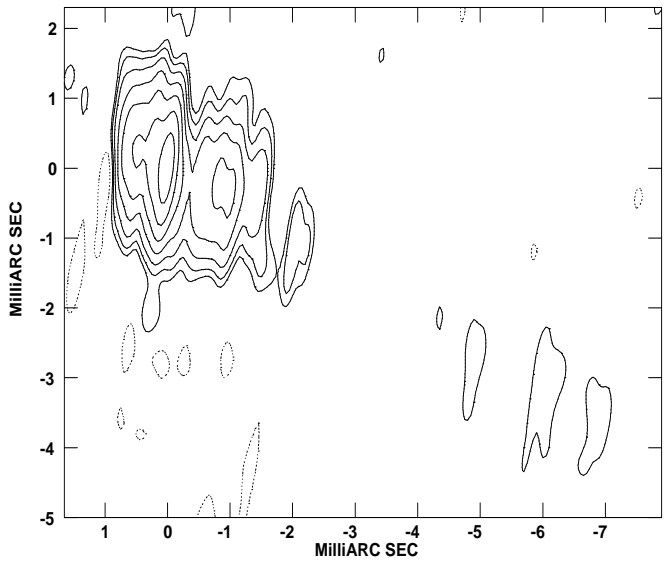,height=8.9cm,width=8.9cm}
\caption[]{ 3C\,273 at 22\,GHz (12Dec92). Contours are -1, 1, 2,
4, 8, 16, 32, 64\,\%. The peak flux density is 5.4 Jy/beam.
Beam FWHM: 1.28 $\times$ 0.19 (mas) at PA -8.5$^\circ$.}
\end{figure}
\begin{figure}
\psfig{figure=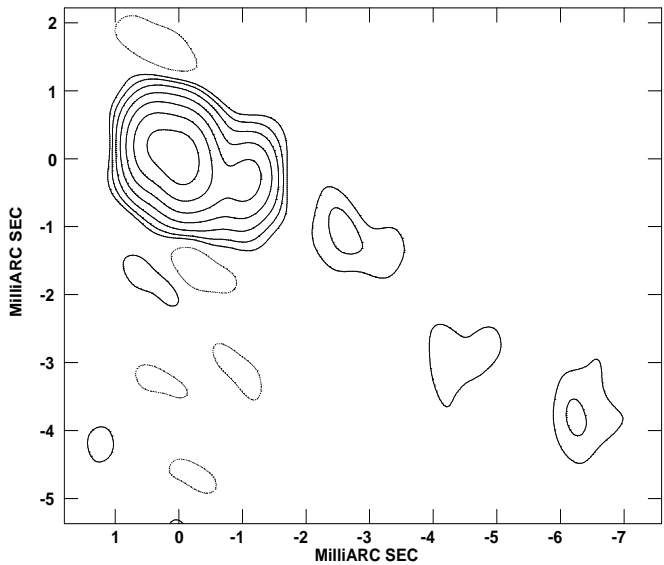,height=8.9cm,width=8.9cm}
\caption[]{ 3C\,273 at 43\,GHz (12Dec92). Contours are -1, 1, 2,
4, 8, 16, 32, 64\,\%. The peak flux density is 4.88 Jy/beam.
Beam FWHM: 0.77 $\times$ 0.53 (mas) at PA 6.8$^\circ$.}
\end{figure}

\begin{figure}
\psfig{figure=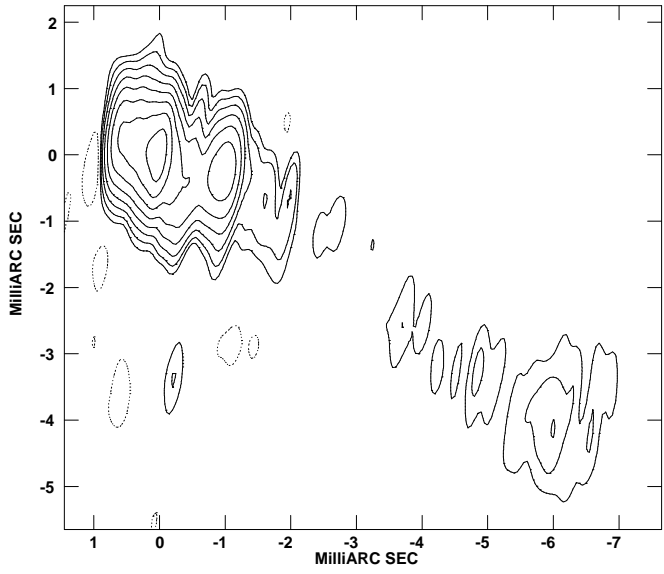,height=8.9cm,width=8.9cm}
\caption[]{ 3C\,273 at 22\,GHz (21Dec92). Contours are -0.5, 0.5, 1, 2,
4, 8, 16, 32, 64\,\%. The peak flux density is 5.9 Jy/beam.
Beam FWHM: 0.99 $\times$ 0.20 (mas) at PA -8.4$^\circ$.}
\end{figure}
\begin{figure}
\psfig{figure=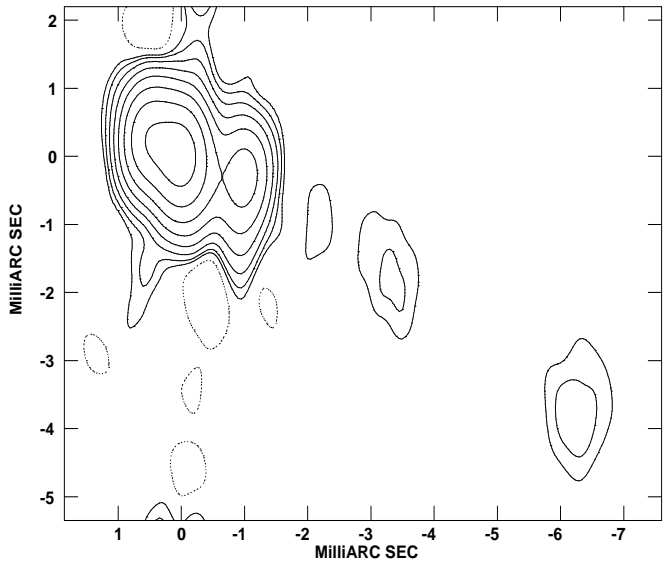,height=8.9cm,width=8.9cm}
\caption[]{ 3C\,273 at 43\,GHz (21Dec92). Contours are -0.5, 0.5, 1, 2
4, 8, 16, 32, 64\,\%. The peak flux density is 3.67 Jy/beam.
Beam FWHM: 0.74 $\times$ 0.35 (mas) at PA -8.9$^\circ$.}
\end{figure}
\begin{figure}
\psfig{figure=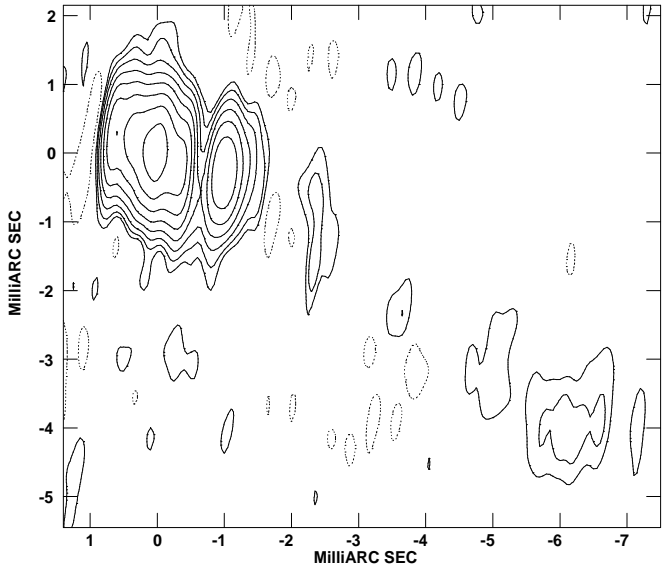,height=8.9cm,width=8.9cm}
\caption[]{ 3C\,273 at 22\,GHz (04Jan93). Contours are -0.5, 0.5, 1, 2,
4, 8, 16, 32, 64\,\%. The peak flux density is 4.19 Jy/beam.
Beam FWHM: 1.03 $\times$ 0.19 (mas) at PA -6.1$^\circ$.}
\end{figure}
\begin{figure}
\psfig{figure=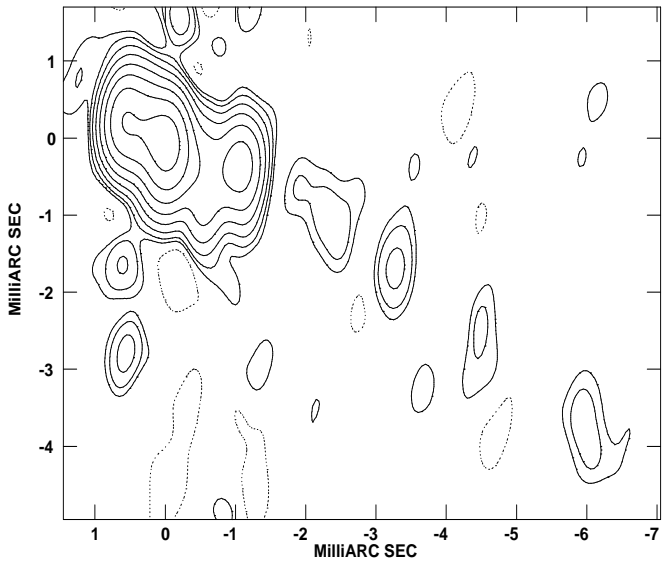,height=8.9cm,width=8.9cm}
\caption[]{ 3C\,273 at 43\,GHz (04jan93). Contours are -0.5, 0.5, 1, 2,
4, 8, 16, 32, 64\,\%. The peak flux density is 2.26 Jy/beam.
Beam FWHM: 0.74 $\times$ 0.36 (mas) at PA -4.6$^\circ$.}
\end{figure}
\begin{figure}
\psfig{figure=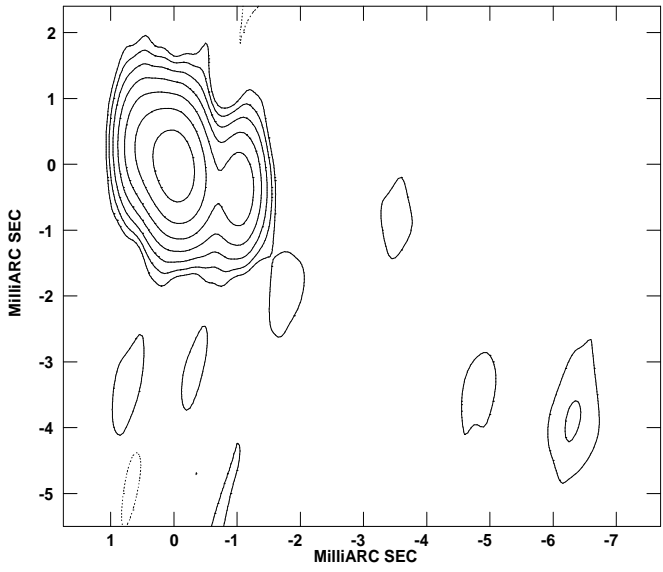,height=8.9cm,width=8.9cm}
\caption[]{ 3C\,273 at 22\,GHz (23Jan93). Contours are -1, 1, 2,
4, 8, 16, 32, 64\,\%. The peak flux density is 5.81 Jy/beam.
Beam FWHM: 1.3 $\times$ 0.5 (mas) at PA 0$^\circ$.}
\end{figure}
\begin{figure}
\psfig{figure=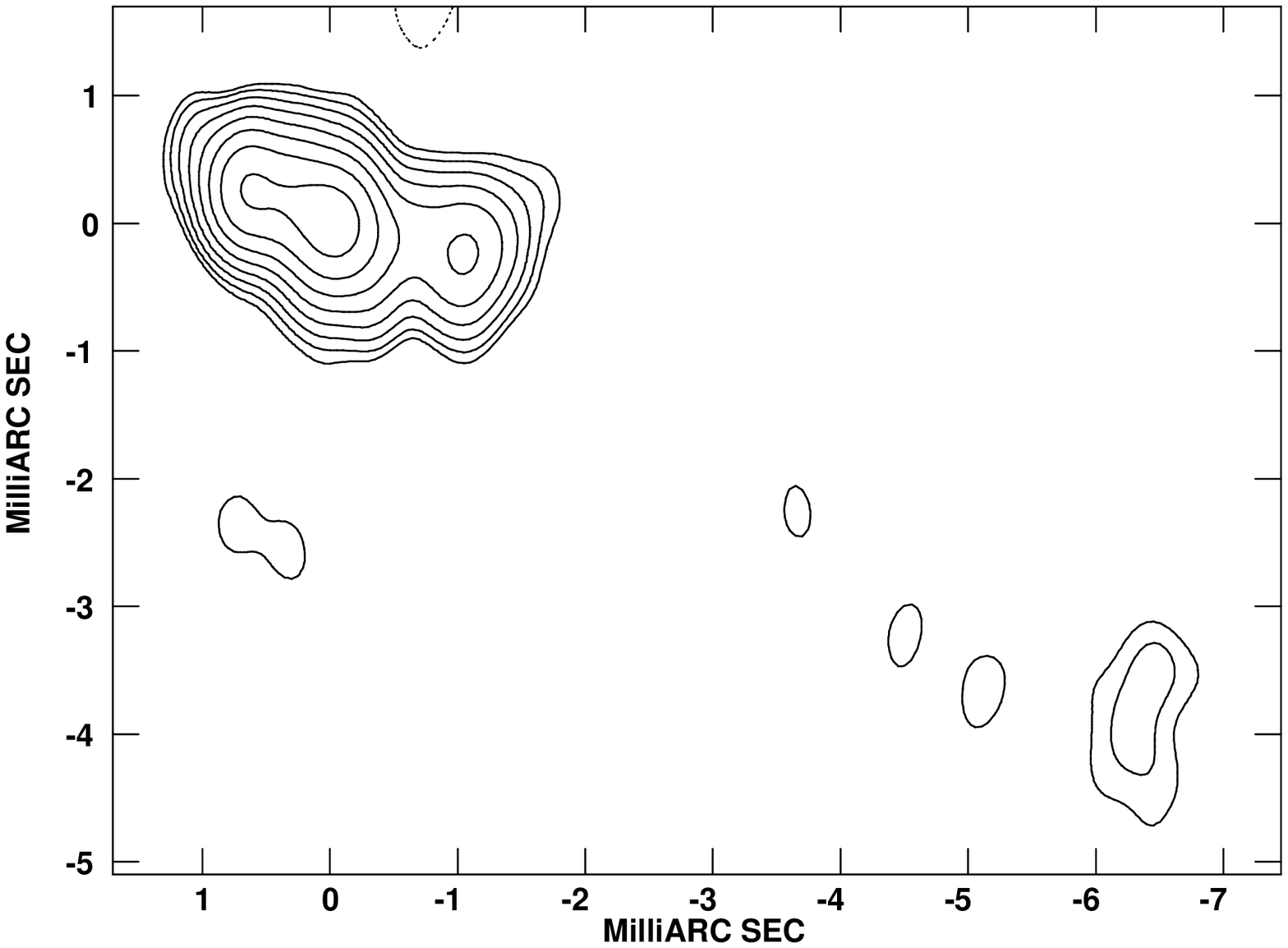,height=8.9cm,width=8.9cm}
\caption[]{ 3C\,273 at 43\,GHz (24Jan93). Contours are -0.5, 0.5, 1, 2,
4, 8, 16, 32, 64\,\%. The peak flux density is 4.11 Jy/beam.
Beam FWHM 0.62 $\times$ 0.37 (mas) at PA 0.6$^\circ$.}
\end{figure}
\subsection{The overall structure}
With the observations at 22\,GHz and 43\,GHz we were able to track the jet 
up to a distance of $\sim$8 mas ($\sim$12 pc) from the core.  We identify
the core with the easternmost component and we furthermore assume this to
be stationary.  The jet consists of a bright region of emission,
which extends up to 2\,mas and a region of much weaker emission
which extends from 2 to 8 mas. The resolution achieved was not enough
to resolve the jet in a direction transverse to the major axis. The images
confirm that the jet is not collinear and has a wiggling structure.

The image from the best available data set at 22\,GHz 
(project bj5b; 21Dec92)
has a beam of 1 $\times$ 0.5 mas and shows that the jet can be tracked 
as far as
$\sim$15 mas from the core (see Fig.\,9). The ridge line of emission clearly 
oscillates around the main orientation. We will discuss this subject in more
detail in section 3.5.

\begin{figure}
\epsfig{figure=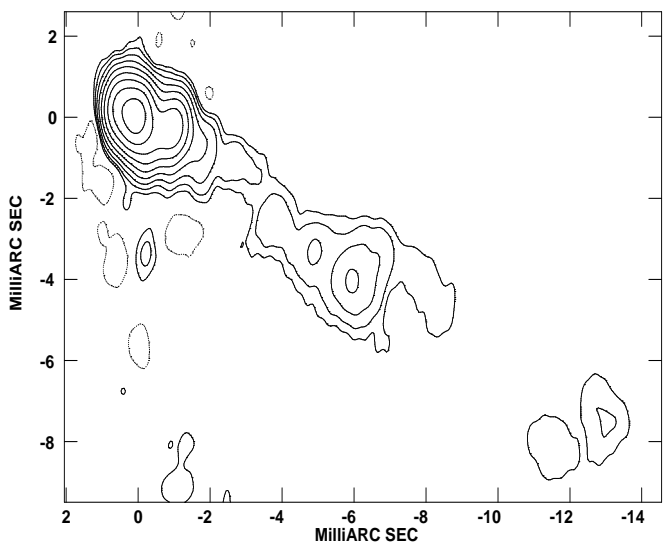,height=8.9cm,width=8.9cm}
\caption[]{ 3C\,273 at 22\,GHz (21Dec92). Contours are -0.25, 0.25, 0.5, 0.5, 
1, 2, 4, 8, 16, 32, 64\,\%. The peak flux density is 9.75 Jy/beam.
Beam FWHM: 1.0 $\times$ 0.5 (mas) at PA -8$^\circ$.}
\end{figure}

The jet major axis has a Position Angle (PA) of $-124^\circ\pm2^\circ$. It
is interesting to compare that PA with those available in the literature
for interferometric observations made at different resolutions.
As shown in Tab.\,2, the PA changes from $-137^\circ$ for the arcsecond
jet, which is aligned to an accuracy of $0.020\arcsec$ with the optical
jet (Bachall et al.\ 1995) to $-120^\circ$ for the $<$1 mas jet.
\begin{table}[h]
\centerline{\bf Table\,2 - The Position Angle of the jet major axis in 3C\,273 }
\vspace{0.5cm}
\hspace{0.5cm} 
\begin{tabular}{rccclrl}
\hline
Freq.    &     & HPBW  &        & Range  & P.A.   & Reference                    \\
GHz      & mas &  mas  &  deg   &  mas   &  deg &                                \\
\hline
0.4      &$10^3$&$10^3$&        &0--25
                            $\times10^3$ & $-137$  & Davis et al.\ 1985          \\
5.0      &  2  &   1   &$-12$   &0--50   & $-130$ & Zensus et al.\ 1988          \\
22.2     & 0.25& 1.9   &$-10$   &0--2    & $-128$  & Zensus et al.\ 1990         \\
22.2     & 0.96& 0.27  & $-9$   &0--8    & $-122$  & Lepp\"anen et al.\ 1995     \\
43.0     & 0.5 & 0.13  & $-7$   &0--0.2  & $-100$  & Krichbaum et al.\ 1990      \\
         &     &       &        &0--0.8  & $-115$  &                             \\
43.0     & 0.8 & 0.4   & $-4$   &0--0.4  & $-100$  & present paper               \\
         &     &       &        &0--0.7  & $-117$  &                             \\
         &     &       &        &0--1    & $-123$  &                             \\
         &     &       &        &0--2    & $-110$  &                             \\
         &     &       &        &0--8    & $-124$  &                             \\
100.0    & 0.05& 0.280 & $-4.4$ &0--0.250& $-119$  & B\aa\aa th et al.\ 1991     \\
         &     &       &        &0--0.080& $-160$  &                             \\
\hline
\end{tabular}
\vspace{0.5cm}
\end{table}

\subsection{The first 2 mas of the jet}
To be able to describe the structure of the source and check
for any structural variations, we have model-fitted with gaussian components 
the final self-calibrated visibilities of each data set. The criteria 
adopted to model the source are that the model should be as 
simple as possible and fit the visibilities with a good 
agreement factor. As one can see from the images in Figs.\,1--8, the source
is dominated by the central 2 mas-jet. The weak
components which lie along the jet, at a radial distance from the core larger 
than 2 mas, only marginally affect the best fit between model and data.
In other words, these components are not well constrained by the data.  
Therefore those components are mainly missing from the Tables. 
Table\,3 lists the models for the 22\,GHz data sets, 
Table\,4 lists the models for the 43\,GHz data sets and
Table\,5 lists the models for the 22\,GHz data sets for baselines
shorter than 450 M$\lambda$. The formal errors associated with the positions of
each component as derived from the modelfitting procedure are rather small. 
 
A more realistic estimate of the errors comes from independent attempts to 
model 
the source.  This gave a distribution of positions for each component from 
which errors have been estimated.   For components which are compact 
compared to the beam size, the errors associated with their positions are
estimated to be $\sim$10\% of the beam. For more extended components,
the associated errors are $\sim$10\% of the full width, half maximum of the
gaussian derived from the modelfitting.   


\begin{table}[h]
\centerline{\bf Table\,3 - Modelfitting 22\,GHz data sets}
\vspace{0.5cm}
\hspace{0.5cm} 
\begin{tabular}{lccrccr}

bj5a    & Flux & Radius & Theta  & Major  & Axial & Phi        \\
12Dec92 & (Jy) & (mas)  & (deg)  & (mas)  & ratio & (deg)      \\
\hline
        & 1.95 &  0.80  &  53.2   &  0.36 & 0.59  &    0.0     \\
        & 3.45 &  0.41  &  49.9   &  0.36 & 0.58  & $-$6.5     \\
        & 5.90 &  0.19  &  30.4   &  0.29 & 0.46  & $-$4.1     \\
        & 1.47 &  0.26  &$-$157.2 &  0.30 & 0.46  &    0.0     \\
        & 2.46 &  0.37  &$-$112.2 &  0.47 & 0.39  & $-$6.4     \\
        & 2.64 &  0.98  &$-$105.4 &  0.48 & 0.45  &    0.0     \\
        & 0.30 &  1.88  &$-$128.2 &  0.27 & 0.51  & $-$2.4     \\
        & 0.93 &  7.11  &$-$115.7 &  1.70 & 0.29  &   40.9     \\
        & 0.14 &  7.30  &$-$22.2  &  0.26 & 0.56  & $-$2.9     \\
\hline
            \\
bj5b    & Flux & Radius & Theta & Major & Axial & Phi      \\
21Dec92 & (Jy) & (mas)  & (deg) & (mas) & ratio & (deg)    \\
\hline
        & 2.62 & 0.63   &  71.1 & 0.32  & 0.11  & 53.1     \\
        & 9.10 & 0.22   &  68.5 & 0.59  & 0.23  & 50.7     \\
        & 3.73 & 0.00   & $-$37.6 & 0.28  & 0.28  &$-$24.1     \\
        & 4.07 & 0.56   &$-$125.3 & 0.78  & 0.89  &$-$61.5     \\
        & 2.24 & 1.03   &$-$105.0 & 0.33  & 0.72  & 39.4     \\
        & 0.23 & 1.79   &$-$127.5 & 0.85  & 0.31  &179.6     \\
        & 0.12 & 2.18   &$-$110.6 & 0.97  & 0.02  &170.7     \\
        & 1.60 & 6.72   &$-$123.1 & 3.16  & 0.73  & 66.5     \\
\hline
             \\
bj5c    & Flux & Radius & Theta & Major & Axial & Phi      \\
04Jan93 & (Jy) & (mas)  & (deg) & (mas) & ratio & (deg)    \\
\hline
        & 1.11 & 0.66   &  78.6 & 0.22  & 0.68  &  68.2    \\
        & 8.26 & 0.14   &  68.9 & 0.92  & 0.20  &  49.4    \\
        & 4.00 & 0.07   &$-$4.8 & 0.27  & 0.18  & 119.8    \\
        & 2.24 & 0.40   &$-$122.9 & 0.40  & 0.57  &   1.1    \\
        & 2.66 & 1.07   &$-$105.8 & 0.37  & 0.65  &  56.8    \\
        & 0.36 & 7.42   &$-$123.5 & 1.04  & 0.59  &  54.5    \\
\hline
            \\
bj5e    & Flux & Radius & Theta & Major & Axial & Phi      \\
23Jan93 & (Jy) & (mas)  & (deg) & (mas) & ratio & (deg)    \\
\hline
        & 1.84 & 0.70   &  63.02& 0.40  & 0.37  &   13.6   \\
        & 3.79 & 0.45   &  57.69& 0.34  & 0.21  &  $-$62.9   \\
        &12.36 & 0.03   & $-$96.32& 0.42  & 0.70  &   55.6   \\
        & 1.65 & 0.46   &$-$122.95& 0.54  & 0.11  &  $-$27.0   \\
        & 3.85 & 1.05   &$-$105.76& 0.52  & 0.59  &  $-$78.9   \\
        & 0.40 & 7.28   &$-$122.79& 1.24  & 0.57  &  $-$14.1   \\
\hline
\end{tabular}
\vspace{0.5cm}
$Note$: components are ordered from East to West 
\end{table}


\begin{table}[h]
\centerline{\bf Table\,4 - Modelfitting 43\,GHz data sets}
\vspace{0.5cm}
\hspace{0.5cm} 
\begin{tabular}{lccrccr}

bj5a    & Flux & Radius & Theta & Major & Axial & Phi      \\
12Dec92 & (Jy) & (mas)  & (deg) & (mas) & ratio & (deg)    \\
\hline
        & 3.24 & 0.50   &  66.9 & 0.33  &  0.46 &   0.0    \\
        & 2.98 & 0.13   &  13.1 & 0.35  &  0.56 &   0.0    \\
        & 4.18 & 0.20   &$-$125.9 & 0.62  &  0.56 &  $-$22.1   \\
        & 0.63 & 0.83   &$-$127.9 & 0.38  &  0.53 &  $-$6.8    \\
        & 1.25 & 1.16   &$-$104.0 & 0.43  &  0.47 &   1.6    \\
        & 0.06 & 2.27   &$-$104.9 & 0.24  &  0.73 &  $-$5.8    \\
        & 0.13 & 2.75   &$-$114.0 & 0.33  &  0.42 &  $-$7.7    \\
        & 0.12 & 3.53   &$-$114.1 & 0.46  &  0.45 &  $-$11.5   \\
        & 0.17 & 7.37   &$-$120.5 & 0.53  &  0.47 &   2.6    \\
\hline
                    \\
bj5b    & Flux & Radius & Theta & Major & Axial & Phi      \\
21Dec92 & (Jy) & (mas)  & (deg) & (mas) & ratio & (deg)    \\
\hline
        & 2.53 & 0.65   &  66.6 & 0.42  & 0.37  &  $-$8.6   \\
        & 2.20 & 0.34   &  50.9 & 0.33  & 0.40  & $-$65.0   \\ 
        & 4.55 & 0.04   &$-$149.0 & 0.46  & 0.48  & $-$19.2   \\ 
        & 1.00 & 0.53   &$-$132.4 & 0.53  & 0.47  &  $-$9.6   \\
        & 1.25 & 1.06   &$-$106.1 & 0.36  & 0.62  &  $-$2.2   \\
        & 0.11 & 4.21   &$-$119.9 & 0.29  & 0.45  &   0.0   \\
        & 0.07 & 7.44   &$-$120.1 & 0.43  & 0.34  &  $-$8.9   \\
\hline
                 \\
bj5c    & Flux & Radius & Theta & Major & Axial & Phi      \\
04Jan93 & (Jy) & (mas)  & (deg) & (mas) & ratio & (deg)    \\
\hline
        & 1.59 & 0.64   &  69.5 & 0.42  & 0.31  &  $-$2.3  \\
        & 0.97 & 0.36   &  63.1 & 0.34  & 0.40  &  $-$6.8  \\
        & 3.42 & 0.10   &$-$165.7 & 0.48  & 0.61  &   3.1  \\
        & 0.40 & 0.72   &$-$134.2 & 0.36  & 0.39  &  $-$8.2  \\
        & 0.70 & 1.11   &$-$110.3 & 0.36  & 0.48  &   0.0  \\
        & 0.09 & 3.52   &$-$117.6 & 0.34  & 0.39  &  $-$6.4  \\
\hline
                     \\   
bj5e    & Flux & Radius & Theta & Major & Axial & Phi      \\
24Jan93 & (Jy) & (mas)  & (deg) & (mas) & ratio & (deg)    \\
\hline
        & 3.03 & 0.71 &  66.0  & 0.20  & 0.94  &  15.9  \\
        & 1.89 & 0.40 &  56.1  & 0.30  & 0.40  & $-$44.1  \\
        & 5.71 & 0.02 & $-$93.8  & 0.34  & 0.89  &  87.5  \\
        & 1.31 & 0.65 &$-$116.6  & 0.70  & 0.58  & $-$19.9  \\
        & 1.04 & 1.15 &$-$102.1  & 0.47  & 0.86  &  $-$6.0  \\
        & 0.23 & 7.44 &$-$121.9  & 0.92  & 0.40  & $-$15.5  \\
\hline
\end{tabular}
\vspace{0.5cm}
$Note$: components are ordered from East to West 
\end{table}


\begin{table}[h]
\centerline{\bf Table\,5 - Modelfitting of the 22\,GHz data sets using 
                          baseline length $\leq 450 M\lambda$}
\vspace{0.5cm}
\hspace{0.5cm} 
\begin{tabular}{lccrccr}

bj5a    & Flux & Radius & Theta & Major & Axial & Phi      \\
12Dec92 & (Jy) & (mas)  & (deg) & (mas) & ratio & (deg)    \\
\hline
        & 2.66 & 0.67   &  53.0 & 0.33  & 0.46  &  65.5  \\
        & 3.51 & 0.33   &  54.7 & 0.37  & 0.56  &  24.1  \\
        & 6.85 & 0.03   &  $-$2.2 & 0.46  & 0.41  & $-$19.3  \\
        & 2.16 & 0.53   &$-$133.2 & 0.38  & 0.53  &  77.9  \\
        & 3.01 & 1.06   &$-$106.7 & 0.44  & 0.47  &  23.5  \\
        & 0.54 & 2.19   &$-$131.1 & 0.56  & 0.52  &  54.8  \\
        & 0.81 & 6.93   &$-$116.7 & 1.26  & 0.55  &  41.6  \\
\hline
               \\
bj5b    & Flux & Radius & Theta & Major & Axial & Phi      \\
21Dec92 & (Jy) & (mas)  & (deg) & (mas) & ratio & (deg)    \\
\hline
        & 1.68 &  0.67  &  73.6 &  0.32 &  0.53 &  $-$42.3  \\
        & 7.02 &  0.34  &  58.5 &  0.53 &  0.37 &   60.7  \\
        & 7.37 &  0.03  & 162.3 &  0.44 &  0.50 &  $-$18.2  \\
        & 3.30 &  0.58  &$-$123.5 &  0.55 &  1.00 &    0.0  \\
        & 2.99 &  1.05  &$-$104.4 &  0.47 &  0.67 &   13.5  \\
        & 0.59 &  7.12  &$-$124.8 &  0.72 &  0.50 &   57.3  \\
\hline
              \\
bj5c    & Flux & Radius & Theta & Major & Axial & Phi      \\
04Jan93 & (Jy) & (mas)  & (deg) & (mas) & ratio & (deg)    \\
\hline
        & 1.84 & 0.65   &  68.4 & 0.34  & 0.63  &  0.0  \\
        & 3.55 & 0.34   &  58.7 & 0.41  & 0.59  &  3.1  \\
        & 7.18 & 0.00   & $-$88.5 & 0.50  & 0.49  &  0.0  \\
        & 2.91 & 0.38   &$-$122.4 & 0.41  & 0.45  &  3.1  \\
        & 3.00 & 1.04   &$-$107.0 & 0.62  & 0.56  &  0.0  \\
        & 0.20 & 2.76   &$-$120.7 & 0.43  & 0.49  & $-$4.5  \\
        & 0.20 & 7.20   &$-$121.6 & 0.47  & 0.44  & $-$4.1  \\
\hline
                \\
bj5e    & Flux & Radius & Theta & Major & Axial & Phi      \\
24Jan93 & (Jy) & (mas)  & (deg) & (mas) & ratio & (deg)    \\
\hline
        & 1.11 & 0.67   &  73.0 &  0.20 &  0.30 &  $-$8.9  \\
        & 3.85 & 0.51   &  53.6 &  0.32 &  0.35 & $-$69.8  \\
        &12.44 & 0.02   &$-$151.2 &  0.40 &  0.63 & $-$86.9  \\
        & 2.65 & 0.52   &$-$129.2 &  0.43 &  0.45 & $-$23.1  \\
        & 3.25 & 1.10   &$-$106.6 &  0.34 &  0.61 & $-$62.1  \\
        & 0.35 & 7.27   &$-$122.7 &  0.83 &  0.67 &  $-$2.9  \\
\hline
\end{tabular}
\vspace{0.5cm}
$Note$: components are ordered from East to West 
\end{table}
The inner part of the jet of 3C\,273 at 43\,GHz can be modelled with five
compact components. The modelfitting was performed using DIFMAP. The
reduced chi-squared, which measures the goodness of the fit, gives values
close to 1 (see Pearson 1994). In order to verify the goodness of the models,
the following checks were done: (1) to verify that the source structure was actually
changing with epochs, each model was used to fit the visibilities
of an other data set; in each case, a poor fit was always obtained; 
(2) each model was used
to make a restored map which was compared by eye with the related hybrid
image; (3) the ratio between the total flux density from the hybrid images
and the flux density obtained adding up the flux density from the individual
components in the related model was always close to 1. This last result also
suggests that the error estimate associated with the flux density of each
component is $\sim$10\%.

Each components position in polar coordinates has been projected along two
orthogonal directions and plotted in Fig.\,10. The components have been
shifted in such a way that the easternmost components, assumed to be the
'core', do overlap. It is clear that the components are not collinear and
that the PA of the line joining the core with each of the components position
changes dramatically, oscillating around a line in PA$\simeq$ $-124^\circ$.

In order to prove if such a structure is real, we have model-fitted the 22\,GHz
data sets using the visibilities from baselines in the range
$\leq 450$ M$\lambda$ to match the resolution to that of the 43\,GHz
observations in the east-west direction. (At 43\,GHz, the resolution was 
poorer in the east-west direction than at 22\,GHz but better in the
north-south direction.)  As a result, a more sparse coverage of the 
{\it uv}-plane was achieved at 22\,GHz.  The image is plotted in Fig.\,11. 
The bends along the jet are 
less evident because of the poorer resolution in north-south of these 
observations, however, the curved structure is confirmed. The variation in
the positions in {\it y}-direction between 22\,GHz and 43\,GHz for each
component might be caused by opacity effects.  However, we are comparing
observations made in the steep part of the spectrum so the displacement could
be caused by the relatively sparse coverage of the {\it uv}-plane at 22\,GHz.
The variation of component position as a function of opacity
at 22\,GHz and 43\,GHz in 4C\,39.25 was been examined by Alberdi et al.\ (1997)
; they find no evidence for any displacement. 
\begin{figure}
\epsfig{figure=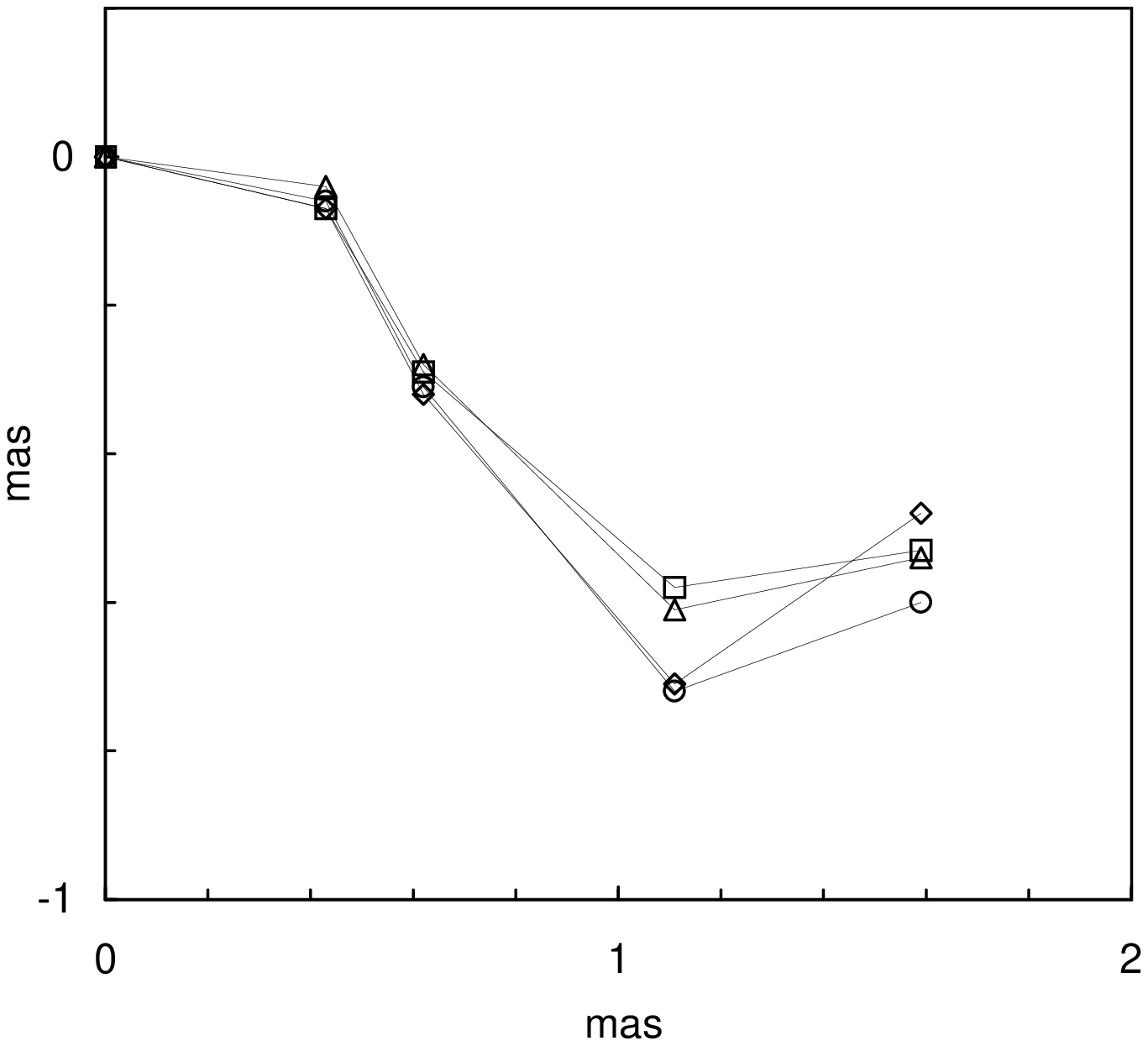,height=12cm,width=12cm}
\caption[]{ The location of the model-fitting components in the inner jet 
of 3C\,273 at 43\,GHz. 
The components positions have been shifted in such
a way that the easternmost components do overlap. 
Symbols: $\diamondsuit$ bj5a, $\triangle$ bj5b, $\circ$ bj5c, $\Box$ bj5e.
The error bars are about the size of the symbols.}
\end{figure}
\begin{figure}
\epsfig{figure=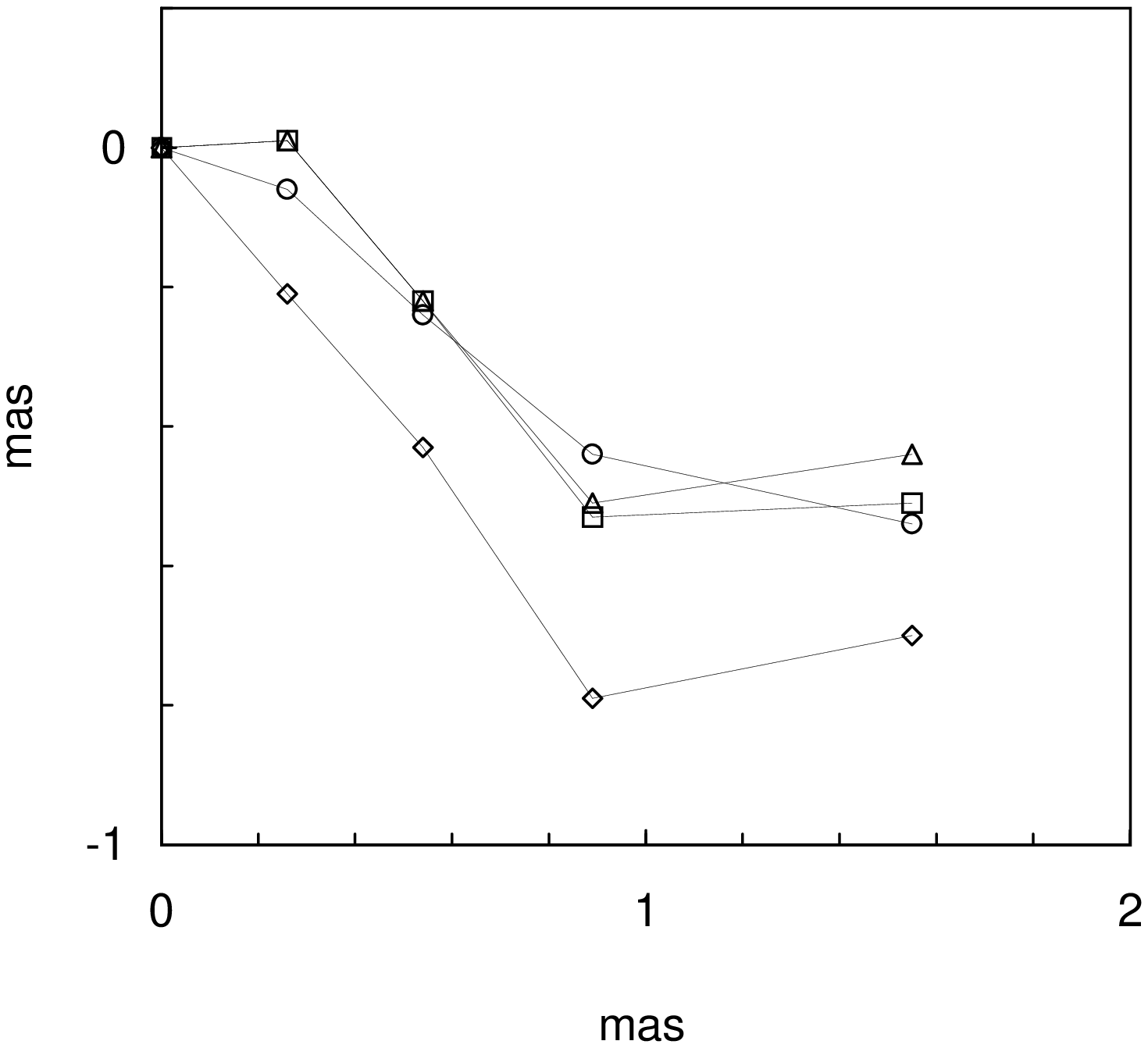,height=12cm,width=12cm}
\caption[]{ The location of the model-fitting components in the inner jet 
of 3C\,273 at 22\,GHz for baselines $<450$M$\lambda$. 
The components positions have been shifted in such
a way that the easternmost components do overlap. 
Symbols: $\diamondsuit$ bj5a, $\triangle$ bj5b, $\circ$ bj5c, $\Box$ bj5e.
The error bars are about the size of the symbols.}
\end{figure}
\subsection{Structural variation with time}
To look for any short term structural variation of 3C\,273, we have plotted
the separation from the core with time of the components along the {\it x} 
and {\it y} directions (East-West and North-South directions respectively) 
projecting the polar vectors of the 43\,GHz models.
 The core is labelled component 1, the next component 2,
than 3, etc... The results are shown in Fig\,12 and Fig\,13 respectively.
Error bars, $\sim$10\% of the HPBW are also plotted. In the plot for the
{\it x}--direction
 the error bars are of the same size of the symbols used. A line, which
represents the best fit, has also been drawn and its correlation coefficient
is reported in Table 6. A value of 1 means the maximum for the
best fit.
\begin{figure}
\epsfig{figure=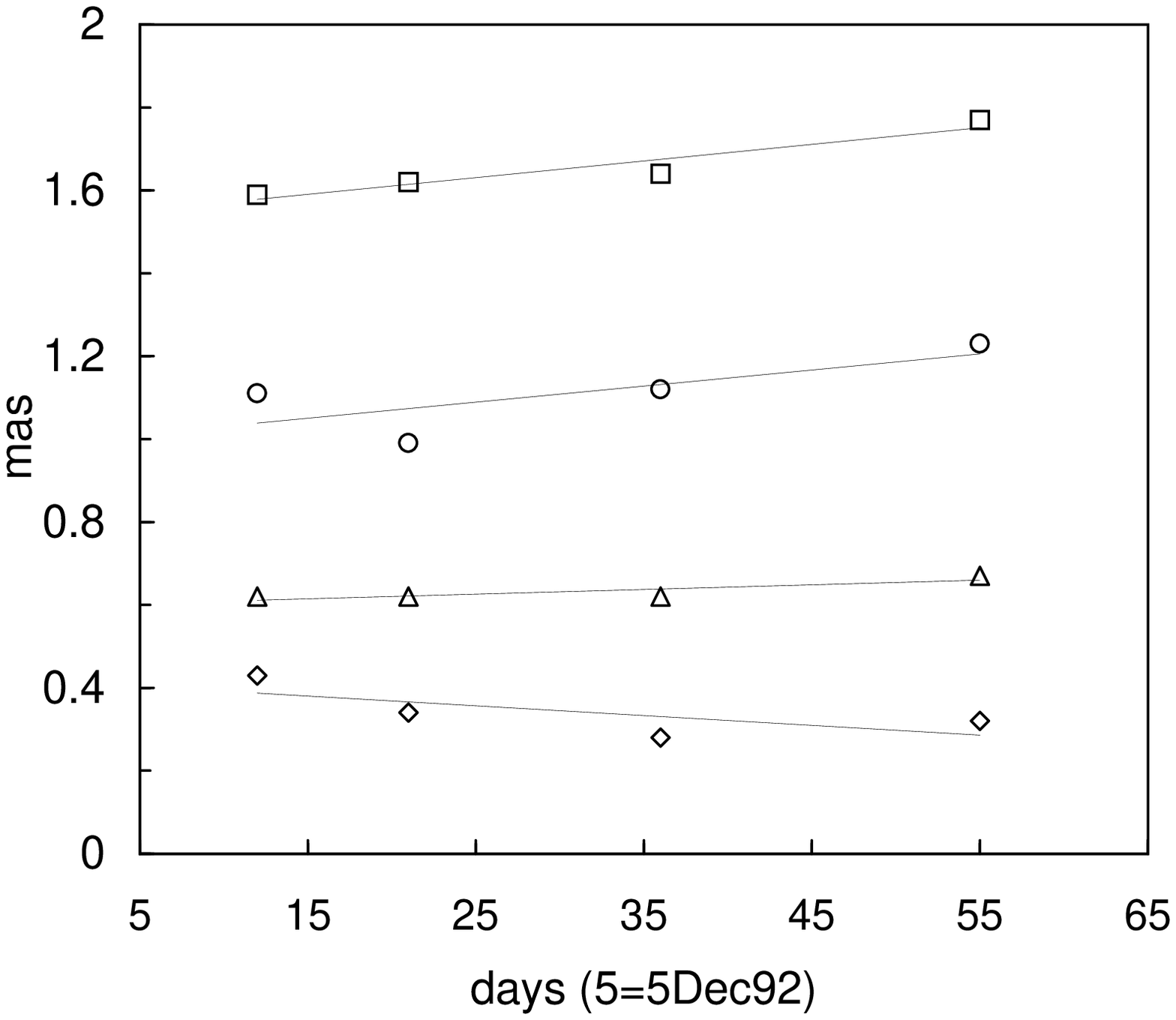,height=12cm,width=12cm}
\caption[]{ The position of the components along the {\it x}--axis {\it vs}
time. The error bars are about the size of the symbol.
Symbols: $\diamondsuit$ comp.\,2, $\triangle$ comp.\,3, $\circ$ 
comp.\,4, $\Box$ comp.\,5.}
\end{figure}
\begin{figure}
\epsfig{figure=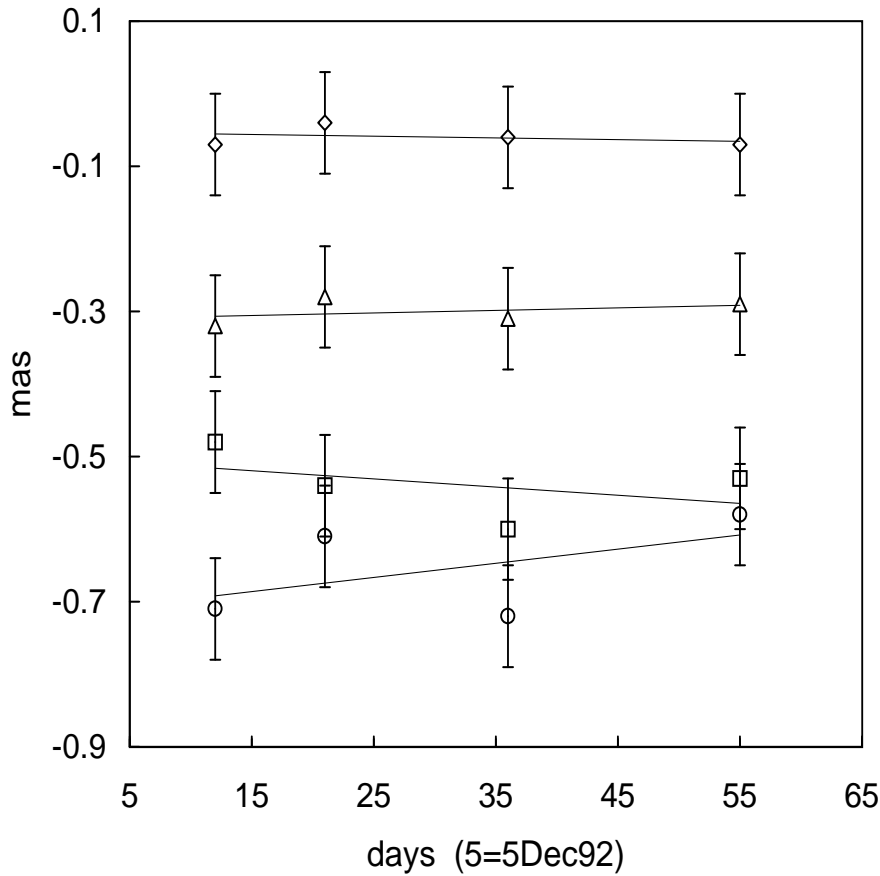,height=12cm,width=12cm}
\caption[]{ The position of the components along the {\it y}--axis {\it vs}
time. Symbols: $\diamondsuit$ comp.\,2, $\triangle$ comp.\,3, $\circ$ 
comp.\,4, $\Box$ comp.\,5.}
\end{figure}
\begin{table}[h]
\centerline{\bf Table\,6 - Apparent speed for modelfitting components in 3C\,273 
at 43\,GHz}
\vspace{0.5cm}
\hspace{0.5cm} 
\begin{tabular}{rrrrrrr}
\hline
Comp.    &{\it x}-axis& corr. &{\it y}-axis& corr. &${\it r}$ & PA      \\
         & mas/yr     & coeff. & mas/yr    & coeff. & mas/yr   & deg   \\
\hline
2        &$-$0.82$\pm$0.09 &$-$0.68 & $<$0.01$\pm$0.08&$-$0.32 &$-$0.82$\pm$0.12 & $\sim90$    \\
3        &  0.42$\pm$0.02 & 0.86 &  0.12$\pm$0.11 & 0.37 &0.44$\pm$0.11& $-$74  \\
4        &  1.39$\pm$0.12 & 0.75 &  0.74$\pm$0.03& 0.54 &1.57$\pm$0.12& $-$62  \\
5        &  1.46$\pm$0.04 & 0.96 &$-$0.40$\pm$0.02 &$-$0.41 &1.51$\pm$0.04& 105   \\
\hline
\end{tabular}
\vspace{0.5cm}
\end{table}
In the {\it x}-direction, components 4 and 5 change their separation from the
core at the same apparent speed, with values in agreement with previous measurements 
obtained with observations which had a much larger time sampling. 
Component 3 has a lower speed, while component 2 shows a separation from the core 
which {\it decreases} with time.   The probability given from the reduced chi-square 
analysis to obtain a similar fit by random is $<25\%$. Such a reverse 
speed, if real, has not been observed in a superluminal source before. 

In the {\it y}-direction, where the positional accuracy is lower,  we have 
sufficient resolution to be able to measure the apparent speed of the
components. Component 2 seems to be stationary with respect to
the core.  Components 3 and 4 are moving north with different speeds, while
component 5 is moving in a direction opposite to component 4. 

In Table\,6 the projected velocities, obtained by combining the
two velocity components, are given (columns 6 and 7 respectively). 
The direction of the radial velocity vectors are not parallel to the jet axis, 
of course, in agreement with the  view that the components in 3C\,273 are moving 
along a wiggling path.
\subsection{Spectral index of individual components}
The observations at 22\,GHz and 43\,GHz were quasi-simultaneous during
the full series of sessions. In order to estimate the spectral indices 
of the components, we have used the
flux densities obtained from model-fitting the source structure. At 22\,GHz,
the adopted flux densities come from  data sets with comparable
{\it uv} ranges to the 43\,GHz observations (i.e.$\leq 450$M$\lambda$).
The spectral 
index of each component for each session is shown in Fig.\,14. 
\begin{figure}
\epsfig{figure=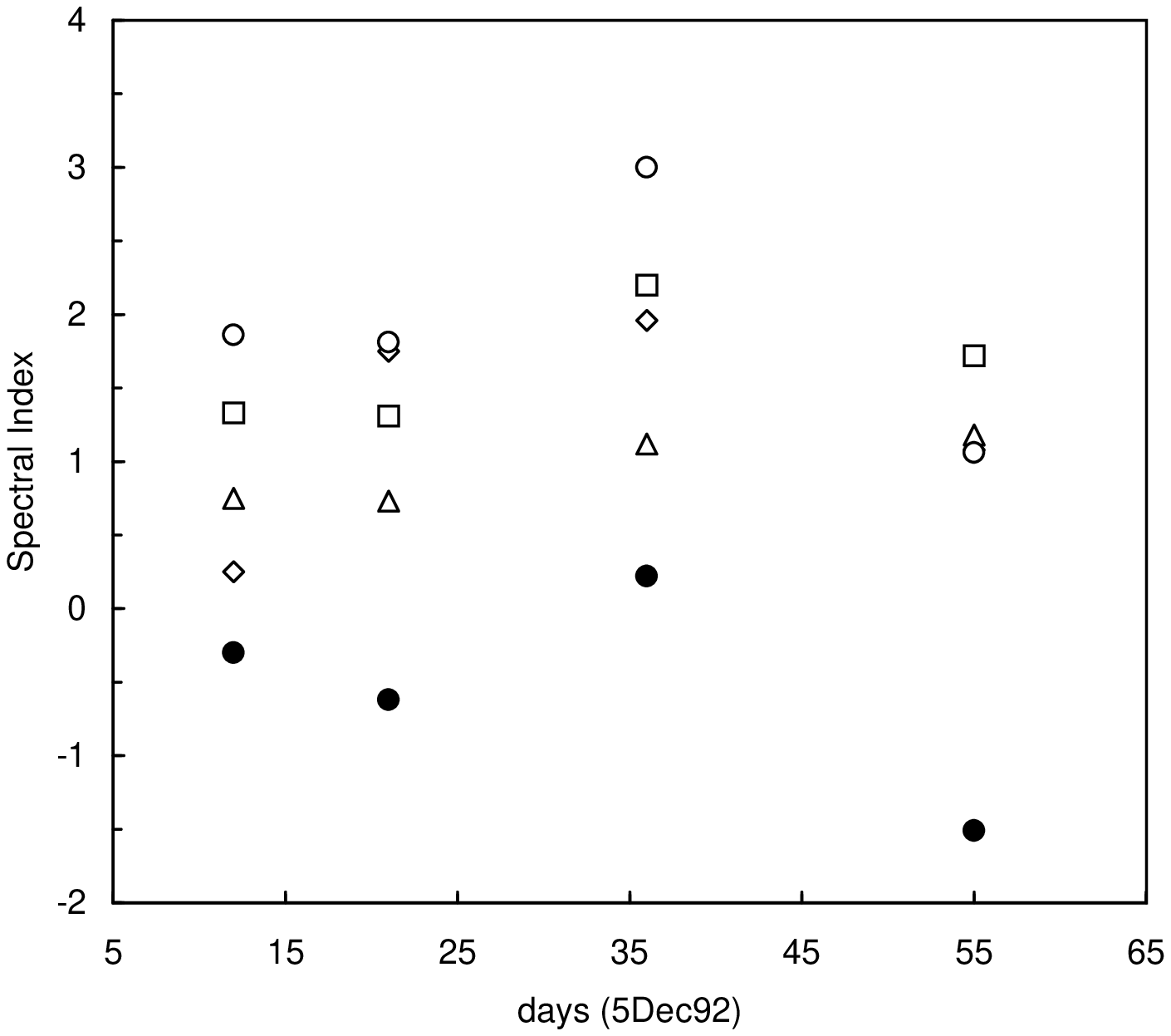,height=12cm,width=12cm}
\caption[]{The spectral indices of the components in the inner jet of 3C\,273 
 versus time.
Symbols: $\bullet$ comp.\,1, $\diamondsuit$ comp.\,2, $\triangle$ comp.\,3, 
$\circ$ comp.\,4, $\Box$ comp.\,5. The component 2 and the component 4 have
the same value of the spectral index on day 55.}
\end{figure}
Among the five components, only component 1, believed to be the core of 
emission, has an inverted spectrum. During the last session, the flux 
density of the core  at 43\,GHz is much higher than at 22\,GHz and the
spectrum more inverted.  This might suggest that a new component is
emerging from the core itself.  All of the components along the jet do show a steep
spectrum, which steepens progressivle with distance from the core. The spectral indices
$\alpha$ ($S\propto\nu^{-\alpha}$), are in
the following ranges: component 1 between 0.2 and $-$1.5;
component 2 between 0.2 and 1.9; component 3 between 0.7 and 1.2;
component 4 between 1.1 and 3.0; component 5 between 1.3 and 2.2.
These spectral indices are quite steep.  This may be because some flux density 
was lost at 43\,GHz because of the better resolution in the
north-south direction.
\subsection{The jet structure in the region 2 -- 8 mas from the core}
It is difficult to model the weak components in the region from 2
to 8 mas from the core.  To visualize the outer 
part of the jet, the position of the peak of the emission for each
blob has been derived from the hybrid maps of Figs.\,1--8.
As a reference point we have taken the strongest component in the
images; this corresponds to component 3 in the model-fitting.

From this we find that, to within the positional accuracy, the components are 
co-moving with
the reference component, {\it i.e.} we do not detect any change in separation with
time between the reference point and any blob in the outer part of jet 
during the period of 
our observations. The position of the peak of emission derived for each 
component from both the 22\,GHz and 43\,GHz images is plotted in Fig.\,15
together with the estimated errors, weighted by the signal-to-noise ratio of 
those components.  The linear-correlation coefficient obtained by fitting 
the points with a straight line gives $-$0.27 which is rather poor.
The straight line drawn in Fig.\,15 represents the major axis 
PA$=-124^{\circ}$ of the jet.  Obviously, the points lie both above and 
below that straight line.  
The second line represents a polynomial, which fits the distribution 
of the series of data  and has a multiple-correlation coefficient very close
to 1. The points at 22\,GHz and 43\,GHz are fully mixed.  Independent 
polynomial fits
to the two series of data give almost identical results. We do not
see any effects of opacity on the component positions within the accuracy
of our measurements.
The jet path is not collinear with the major axis and 
the wiggles, are now clearly tracked.  From Fig.\,15 we can derive the 
wavelength of the oscillation, which is $\approx$6 mas (or $\approx$11 pc).

\begin{figure}
\epsfig{figure=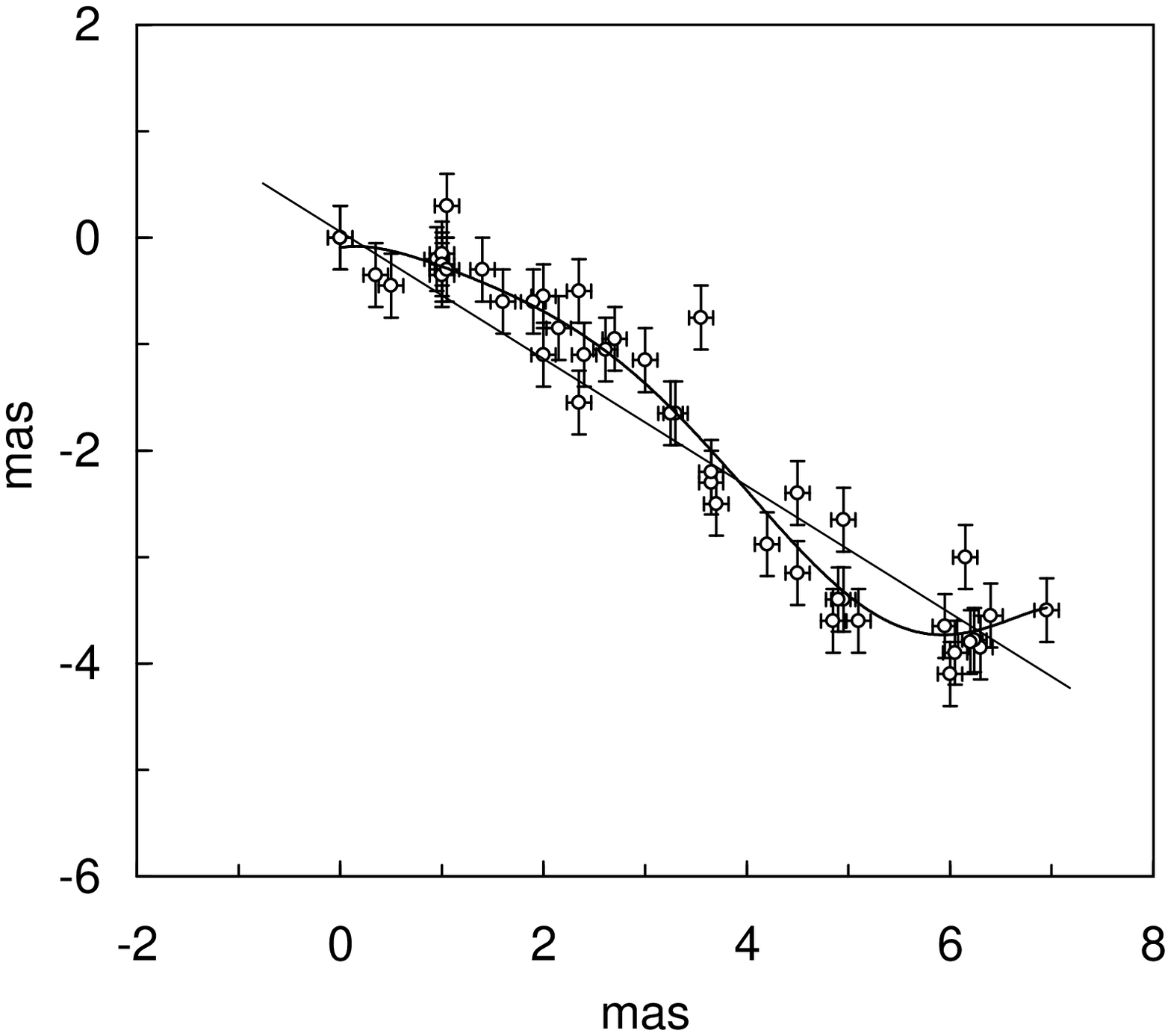,height=12cm,width=12cm}
\caption[]{The position of the peak of emission for the components along
the jet in the range 2--8 mas at 22\,GHz and 43\,GHz.}
\end{figure}
\section{Discussion}
The sub-arcsecond scale structure of the radio jet in 3C\,273 has been 
investigated by many authors using radio interferometric observations. 
In addition to the superluminal motion detected in the core with  VLBI at
high frequencies, these investigations  also show that the 
ridge-line of the jet shows a 'wiggle'.  Davis et al.\ (1985) were able to track 
the ridge over the whole 21 arcseconds length of the jet in their 1 arcsecond 
resolution MERLIN 408\,MHz image.  They interpret the wiggle in terms of precession
in the central object with a period of 15$\times10^3$ years, assuming
a deceleration in the jet material and an apparent tranverse velocity which
equals the superluminal value observed a few mas from the core.     

Zensus et al.\ (1990), combining their 5\,GHz VLBI observations with
those available in literature, pointed out that the ridge line appears to 
oscillate around the main orientation of the jet, with major bends apparently 
being located at about 0.2, 3--4, 10--11 and 15--17 mas from the core.
With their investigation Zensus et al.\ (1990) were also able to track the
components loci in one case, suggesting that that component had
followed the same path.

Similar results were obtained by Krichbaum et al.\ (1990) 
when adding observations at 43\,GHz. They were able to track the bending in
the inner jet down to $\sim$0.5 mas from the core and pointed out that
strongly-bent jets in regions close to the core are a common phenomenon
--- for example, these are seen in 3C\,345, 3C\,84 and 1803+784.

The image of 3C\,273 produced by B\aa\aa th et al.\ (1991) at 100\,GHz with a 
resolution of 50 $\mu$as shows that the jet has a position angle which is
significantly different from those observed at lower frequencies. They
claim that the inner structure has a more pronounced wiggling structure than 
seen on larger scales.

Table\,2 summarizes the PAs of the jet major axis in 3C\,273
at several angular scales. The differences between PAs suggest that the
wiggle can be seen on all angular scales; this suggests that there is 
a common mechanism responsible for this.  From this point of view, the 
observations presented here add important pieces of
information on the inner jet structure of 3C\,273. First of all, in 
the 42 days of monitoring done, we do not
detect any dramatic structural variation along the jet. 
There is a good evidence for jet bending, which starts inside the first
parsec from the core.  We have also been able to measure the apparent 
speeds of the inner jet components in orthogonal directions.
The values obtained are generally consistent
with those available in the literature (i.e. Zensus et al.\ 1990).  
Additionally, we find (i) a counter-moving component  
and (ii) that along the {\it y}--axis, there are components which move 
in opposite direction.
 
Those behaviours are strong indications for a precessing jet. In fact, the 
geometry which permits us to see apparent recessional motion in an 
approaching object 
requires the object to be moving along a spiral path about an axis close
to the observer's line of sight.  Moreover, the tranverse speeds for the outer
components are likely to be higher than for the components closest to the
core.

Alternatively, it is possible that the `core' of the source is
actually component 2.  In this case, components 3, 4 and 5 are moving at lower 
speeds with respect to the core.  Component 1 would then be part of the 
counter-jet.
This does not agree with the observation that component 1 is the only
one showing an inverted spectral index.  Moreover, its apparent speed would be 
significantly lower than that infreed from these observations

The case for  a precessing jet is made stronger by
the distribution of the loci of the peaks of emission for the components
detected in the region of the jet from 2 to 8 mas. Their quasi-sinusoidal
distribution indicates a pattern wavelength of $\approx \lambda\sim$6\,mas.
Assuming an apparent tranverse velocity of $5c$ and using 
\begin{equation}
T = {\lambda \over {c \beta_{obs}}}
\end{equation}  
(Davis et al.\ 1985), we derive a period T$\simeq$7 years. This value should 
be considered as an
upper limit. It is clear from the model-fitting that the jet starts wiggling
even closer to the core. The available resolution is not good enough
to derive a firm period for that oscillation.

The derived period is very short compared with that estimated by Davis 
et al.\ (1985) and two orders of magnitude shorter than that given 
by Roos (1988) for active galaxies during their final, most luminous stage.
A comparably small value of 22 years was derived for the precessing period 
of the jet in 3C279 (Abraham et al.\ 1998). 

It is also worth noting that the positions derived from the 22\,GHz and 43\,GHz
images are fully mixed. The fits for the two series of dots are rather 
similar. Since we are observing in the optically thin part of the spectrum, we
would not see any frequency dependence of positions transverse
to the jet.  This effect would cause the jet to appear more curved at higher
frequency due to opacity effects suggested by Zensus et al.\ (1990).
Observations at lower frequencies (in the optically-thick regime) with similar 
resolution are needed to confirm this definitively.  

The source 3C\,273 is not the only blazar showing curved trajectories.
Sources like 1803$+$784 (Kellermann et al.\ 1998), 3C\,279 
(Unwin et al.\ 1998), 
3C\,84 (Dawhan et al.\ 1998), 3C\,454.3 (Pauliny-Toth 1998), 
0836$+$710 (Otterbein et al.\ 1998) and 3C\,345 (Lobanov \& Zensus 1998) do
show oscillations in their jet trajectories.  The morphology of the last two
sources has been explained in terms of Kelvin-Helmholtz instabilities.
In 3C\,345, the components propagate with constant apparent speed along a 
helically-twisted trajectory.
Hardee (1987) was able to explain this successfully using a model for the 
helical 
twisting of a light relativistic jet expanding in response to an external 
pressure gradient. A similar approch
has been used by  Otterbein et al.\ (1998) to explain the displacements in the
jet of  0836$+$710.  The present observations cannot exclude the possibility 
that the
morphology in 3C\,273 can be modelled with Kelvin-Helmholtz instabilities.

An alternative model  for the helical path of some relativistic jets,
invokes a shock wave propagating along curved trajectories. Such a model has 
been 
discussed by Gomez et al.\ (1993, 1994).  In it, the strongest emission
is obtained when the shock wave reaches the bent regions towards the observer.
\acknowledgements
We thank the staffs at the telescopes and at the correlator
for their kind collaboration. The National Radio Astronomy Observatory is
a facility of the National Science Foundation operated under the cooperative
agreement by Associated Universities Inc.
The referee is thanked warmly for his many constructive comments.  
%
%
%

%

\begin{thebibliography}{}  

\bibitem[]{} Abraham, Z., Carrara, E.A. and Zensus, J.A.: 1998,
IAU Colloquium 164: {\em Radio Emission from Galactic and Extragalactic 
Compact Sources, 1998, APS Conference Series, Vol. 144, pag. 47}, Zensus, A., 
Taylor, G.B. and Wrobel, J.M. eds.
\bibitem[]{} Alberdi, A., Krichbaum, T.P., Graham, D.A. {\it et al.} 1997,
 A\&A 327,513
\bibitem[]{} B\aa\aa th, L.B., Padin, S., Woody, D. {\it et al.} 1991,
 A\&A 241, L1
\bibitem[]{} Bahcall, J.N., Kirhakos, S., Schneider, D.P. {\it et al.} 1995,
APJ 452, L91
\bibitem[]{} Davis, R.J., Muxlow, T.W.B. \& Conway, R.G. 1985, Nature 318, 343
\bibitem[]{} Dhawan, V., Kellermann, K.I. and Romney, J.D.: 1998,
IAU Colloquium 164: {\em Radio Emission from Galactic and Extragalactic 
Compact Sources, 1998, APS Conference Series, Vol. 144, pag. 79}, Zensus, 
A., Taylor, G.B. and Wrobel, J.M. eds.
\bibitem[]{} Gomez, J.L., Alberdi, A and Marcaide, J.M. 1993,  A\&A 274,55
\bibitem[]{} Gomez, J.L., Alberdi, A and Marcaide, J.M. 1994,  A\&A 284,51
\bibitem[]{} Hardee, P.E., 1987, APJ 318,78  
\bibitem[]{} Kellermann, K.I., Vermeulen, R.C., Zensus, A. and Choen, M.H.
1998 AJ, {\it in press}
\bibitem[]{} Krichbaum, T.P., Booth, R.S., Kus, A.J. {\it et al.} 
  1990, A\&A 273, 3
\bibitem[]{} Leach, C.M., McHardy, I.M. \& Papadakis, I.E. 1995, MNRAS 272, 221
\bibitem[]{} Lepp\"anen, K.J., Zensus, J.A. \& Diamond, P.J. 
 1995, AJ 110, 2479.
\bibitem[]{} Lobanov, A.P. and Zensus, J.A. 1998 APJ, {\it in press}
\bibitem[]{} M$^c$Hardy, I.M., Papadakis, I., Leach, C.M. {\it et al.} 
  1994, in {\em IAU Symposium 159}, T.Courvoisier and Blecha eds., pag. 193
\bibitem[]{} Otterbein, K., Krichbaum, T.P., Kraus, A. and Witzel, A.: 1998,
IAU Colloquium 164: {\em Radio Emission from Galactic and Extragalactic 
Compact Sources, 1998, APS Conference Series, Vol. 144, pag. 73},
Zensus, A., Taylor, G.B. and Wrobel, J.M. eds.
\bibitem[]{} Pauliny-Toth, I.I.K.: 1998,
IAU Colloquium 164: {\em Radio Emission from Galactic and Extragalactic 
Compact Sources, 1998, APS Conference Series, Vol. 144, pag. 75}, Zensus, A., 
Taylor, G.B. and Wrobel, J.M. eds.
\bibitem[]{} Pearson T.J.: 1994, {\em Very Long Baseline Interferometry and
the VLBA, 1994, APS Conference Series, Vol. 82, pag. 267}, Zensus, A.,
Diamond, P.J. and Napier, P.J. eds. 
\bibitem[]{} Ross, N.: 1988, Astrophys.Journal 334, 95  
\bibitem[]{} Schmidt, M. 1963, Nature 197, 1040
\bibitem[]{} Shepherd, M.C., Pearson, T.J. \& Taylor, G.B. 1995, BAAS 26, 987
\bibitem[]{} Tornikoski, M., Valtaoja, E., Ter\"asranta, H. {\it et al.} 1994,
 A\&A 289,673
\bibitem[]{} Unwin, S.C., Wehrle, A.E., Xu, W. et al.: 1998,
IAU Colloquium 164: {\em Radio Emission from Galactic and Extragalactic 
Compact Sources, 1998, APS Conference Series, Vol. 144, pag. 69}, Zensus, 
J.A., Taylor, G.B. and Wrobel, J.M. eds.
\bibitem[]{} von Montigny, C., Aller, H., Aller, M., {\it et al.} 1997 483, 161
\bibitem[]{} Zensus, J.A., Unwin, S.C., Cohen, M.H. \& Biretta, J.A.: 1990,
 AJ 100, 1777

\end{thebibliography}
\end{document}